\documentclass{vgtc}                          

\graphicspath{{figures/}{pictures/}{images/}{./}} 

\usepackage{times}                     

\usepackage{tabu}                      
\usepackage{booktabs}                  
\usepackage{lipsum}                    
\usepackage{mwe}                       
\usepackage{balance}
\usepackage{cite}                      
\usepackage{mathptmx}                  

\onlineid{0}

\vgtccategory{Research}

\vgtcinsertpkg

\title{Reimagining Data Visualization to Address Sustainability Goals}

\author{Narges Mahyar\thanks{e-mail: nmahyar@cs.umass.edu}\\ %
        \scriptsize Manning College of Information and Computer Sciences, University of Massachusetts Amherst %
}

\abstract{Information visualization holds significant potential to support sustainability goals such as environmental stewardship, and climate resilience by transforming complex data into accessible visual formats that enhance public understanding of complex climate change data and drive actionable insights. While the field has predominantly focused on analytical orientation of visualization, challenging traditional visualization techniques and goals, through ``critical visualization'' research expands existing assumptions and conventions in the field. In this paper, I explore how reimagining overlooked aspects of data visualization—such as engagement, emotional resonance, communication, and community empowerment—can contribute to achieving sustainability objectives. I argue that by focusing on inclusive data visualization that promotes clarity, understandability, and public participation, we can make complex data more relatable and actionable, fostering broader connections and mobilizing collective action on critical issues like climate change. Moreover, I discuss the role of emotional receptivity in environmental data communication, stressing the need for visualizations that respect diverse cultural perspectives and emotional responses to achieve impactful outcomes. Drawing on insights from a decade of research in public participation and community engagement, I aim to highlight how data visualization can democratize data access and increase public involvement in order to contribute to a more sustainable and resilient future.}


\begin{document}



\firstsection{Introduction}
\maketitle
Visualization has demonstrated immense potential in effectively communicating climate change data to experts, scientists, and policymakers. Given the complexity and interconnected nature of climate change issues, engaging the broader community to collaborate with policymakers and government officials is crucial. However, public engagement with climate change remains low. The pressing question now is: how can we reimagine visualization to effectively communicate this intricate and complex societal challenge to the public?

To envision the future of information visualization for addressing sustainability challenges, we must first look into the past, its history and origins. Understanding the goals and objectives that have shaped the field over time provides valuable insights into its evolution and the foundational principles that guide its development. 
This historical perspective helps to identify areas where there have been limitations or opportunities for growth. By reflecting on these insights, we can reimagine the future of information visualization, considering what needs to be changed or expanded upon to meet new challenges and opportunities in our increasingly data-driven world, especially in the face of sustainability challenges.

Information visualization is inherently interdisciplinary, drawing principles from fields such as statistics, visual communication, graphic design, and cognitive science. This interdisciplinary nature has led to a diverse array of definitions and perspectives. In the following sections, I present notable definitions arranged chronologically to illustrate the evolution of the field, as well as the variety of perspectives and approaches in visualization.

Edward Tufte, in \texttt{The Visual Display of Quantitative Information} \cite{tufte1983visual} 
defines information visualization as a means to visually represent data with the goal of revealing complex information clearly and efficiently. He emphasizes achieving graphical excellence in information visualization. According to Tufte, graphical excellence involves ``the well-designed presentation of interesting data—a matter of substance, statistics, and design'' \cite{tufte1983visual}.
Card et al define visualization as ``the use of computer-supported, interactive visual representations of data to amplify cognition'' \cite{card1999readings}. 
Stephen Few, in \texttt{Information Dashboard Design} \cite{few2006information}, 
defines data visualization as the use of visual representations to explore, comprehend, and communicate data. According to Few, data visualization serves three fundamental goals suited to their respective tasks: exploration for discovery, sensemaking for understanding, and communication for informing decisions. He emphasizes the effectiveness of visual representation in conveying information and highlights vision as the dominant human sense for processing data. 
David McCandless, in \texttt{Information is Beautiful} \cite{mccandless2012information}, views visualization as an art form that transforms data into compelling narratives. 
Tamara Munzner, in \texttt{Visualization Analysis and Design} \cite{munzner2014visualization}, defines visualization as 
``computer-based visualization systems provide visual representations of datasets designed to help people carry out tasks more effectively.''  She emphasizes that visualization is particularly valuable when it complements human capabilities rather than aiming to replace them with automated decision-making processes.
Andy Kirk, in \texttt{Data Visualization: A Handbook for Data Driven Design} 
 defines data visualization as ``the representation and presentation of data to facilitate understanding." He explains three stages of understanding: perceiving, interpreting, and comprehending. 
Colin Ware, in \texttt{Information Visualization: Perception for Design} \cite{ware2019information}, discusses how visual representations leverage human perceptual abilities to recognize patterns and trends. He states that ``one of the greatest benefits of data visualization is the sheer quantity of information that can be rapidly interpreted if it is presented well.''

These diverse definitions reinforce the multifaceted nature of information visualization, encompassing principles from design, cognition, perception, and HCI, and highlighting its role in making data accessible and interpretable. 
While these definitions cover a broad range of goals for information visualization, including discovering new insights, understanding and making sense of data, and communicating data and knowledge through graphical means, the majority of work in the field has focused predominantly on investigating and harnessing analytical aspects of visualization \cite{baumer2022course}. Consequently, less attention has been paid to other societal aspects that consider the lifecycle of a visualization, including its creation, usage by people, and the context in which it is applied.

Drawing on insights gained from over a decade of research in designing and developing visualization for public participation and community engagement using theories of public participation and citizen sourcing \cite{mahyar2016ud, mahyar2018communitycrit, mahyar2019civic, jasim2021communityclick, jasim2021communitypulse, jasim2022supporting, burns2021designing, burns2023we, baumer2022course, aragon2021risingemotions, burns2020evaluate, yavo2023building, mahyar2020designing, reynante2021framework}, I discuss the dimensions in data visualization that can lead to democratizing access to information and fostering increased public engagement. I begin by reviewing recent critical perspectives on visualization that challenge conventional assumptions and delve into broader aspects of the field. Subsequently, I highlight dimensions that require further development or greater emphasis to effectively engage communities with complex data, such as climate change and sustainability information.

By exploring the intersection of data visualization with public participation and participatory frameworks, this work aims to highlight how visual representations of complex information can engage communities, empower them to influence policies, and amplify their voices in decision-making processes. 
This work calls for the development of new theories and approaches in designing visualizations that prioritize Community engagement, and conducting experiments to interrogate previous assumptions. 
Deriving insights from a body of interdisciplinary research, public participation, citizen sourcing, learning and communication theories, the design and development of these visualizations should be methodically implemented and evaluated with target audiences in mind. Future approaches should prioritize clear communication with emotional resonance, stimulate attention, and bridge gaps in understanding to promote inclusivity and facilitate informed dialogues on critical issues such as sustainability and climate change. By emphasizing these aspects, visualizations can not only present data effectively but also resonate emotionally with audiences, driving engagement and prompting actionable outcomes.


\section{Emerging Theories and Critical Developments in Visualization}

More recently, researchers have begun to explore alternative perspectives on visualization, examining the intentions behind designing visualization and its broader implications. These studies offer new perspectives to understanding visualization processes and practices, including ethical dimensions~\cite{correll2019ethical}, feminist perspectives~\cite{d2023data}, critical theory~\cite{dork2013critical}, and rhetorical approaches~\cite{hullman2011visualization}.

In a recent book on \texttt{data visualization in society} \cite{engebretsen2020data}, Alberto Cairo writes a foreword called ``the dawn of a philosophy of visualization'' in which he argues that just as artifacts such as the clock, the compass, or the map has transformed science, society, and culture, data visualization transforms how we perceive and interact with reality. Cairo invites visualization researchers to critically examine data visualization, exploring its origins and intentions, and to work towards developing a philosophy of data visualization. Data Visualization in Society is a notable example of such effort, offering a deeper inquiry into the role of visualization in society and politics and its potential to bridge or deepen societal divides.
This book has inspired much research, including our work on the political aspects of text visualization in the context of civics \cite{baumer2022course}. However, if we examine the citations of this book, we find that it has been mostly applied in other disciplines such as journalism, communication, education, and politics. There is a need for more of such critical work to be incorporated into the core body of visualization research.


In \texttt{Data Feminism}, D'Ignazio and Klein, presents a compelling framework that combines data science and intersectional feminist theory to address issues of power and justice \cite{d2023data}. The book is structured around seven key principles: examining power, challenging power, elevating emotion and embodiment, rethinking binaries and hierarchies, embracing pluralism, considering context, and making labor visible. These principles guide readers in understanding how power dynamics influence data collection, analysis, and interpretation, and how these processes can be redesigned to promote equity and justice. They advocate for a critical approach to data that recognizes these existing social hierarchies and biases and works to dismantle them. For example, they highlight how the male/female binary in data classification can be expanded to challenge other hierarchical systems.
This book has swiftly influenced various disciplines, including visualization. For instance, we have utilized concepts from this book to lay the foundation for the integration of para data into visualization \cite{burns2020evaluate}, advocate for including novices into the field, and critically examining how power dynamics shape the definition of valuable and valid knowledge or skills \cite{burns2023we}, and emphasizing that data is neither neutral nor objective which is extremely important in political contexts such as civic decision-making \cite{baumer2022course}. 

While these works have paved the way for new and critical perspectives in data visualization, we need more research that critically investigates which aspects, goals, and methods are most effective in engaging a global audience with climate change data. There are a wide range of open questions to consider: 
What adjustments are required in our visualization design, methods, processes, and evaluation when engagement becomes the primary goal of data visualization? How can we engage, inform, and inspire such a diverse audience not only to understand the urgency of climate data but also to take meaningful action? How can embedding emotion and embodiment enhance the communication of complex data to the general public? How can we make data more relatable, personal, and impactful? How do we empower audiences with varying levels of data, language, and visual literacy to understand the data and enhance their analytical reasoning? Should we explore new media? Should we incorporate multimodal interactions? Should we integrate art, engage with the various senses such as smell and sound in new ways? What is the role of sonification,  music and theatrical performances in this context? 

Viegas and Wattenberg \cite{viegas2011make} emphasized that ``an ideal visualization should not only communicate clearly but also stimulate viewer engagement and attention'', which is particularly relevant in this context. However, recent research by Mark 
reveals a significant decline in average attention spans on screens, dropping from around two and a half minutes in 2004 to approximately 75 seconds \cite{mark2023attention}. According to Mark, currently, people can only maintain focus on one screen for an average of 47 seconds. Therefore, the pressing question remains: in an era of shrinking attention spans, how do we capture and sustain engagement?

Furthermore, how can we empower stakeholders across diverse sectors to grasp climate data and effectively address climate-related challenges? 
How can visualization not only drive progress in climate mitigation but also lay the foundation for a more resilient and sustainable world?

\section{Fostering Community Engagement: Dimensions and Theoretical Development}

Designing technology solutions for sociotechnical problems presents significant challenges due to the interconnected nature of these issues and the complexity arising from diverse stakeholder needs, values, objectives, knowledge levels, and technical skills \cite{mahyar2020designing}. In the literature of Computer-Supported Cooperative Work (CSCW), it is well-known that transitioning from supporting a single user to multiple users adds layers of complexity \cite{isenberg2009collaborative, mahyar2014supporting}. While visualization design has a strong foundation in supporting groups in CSCW research, an important distinction lies in designing for heterogeneous groups with diverse goals and objectives, rather than homogeneous groups of collaborators with shared interests and similar skills. 

When designing for communities, particularly in the context of sociotechnical problems, the challenge lies in accommodating varying levels of knowledge, technical backgrounds, visual and data literacy, objectives, values, available time, attention, and more. Complicating this space further are inherent conflicts, such as the NIMBY (Not In My Backyard) problem in urban planning or misinformation and disbelief in climate change. Addressing these challenges necessitates nuanced approaches that promote inclusivity, mitigate conflicts, and enhance community engagement through effective communication and visual strategies.

Therefore, there is a crucial need for engaging, accessible, and relatable data visualization for people with various capabilities. Despite advancements in the field of visualization, current solutions often fall short, lacking inclusivity and accessibility for novices \cite{burns2023we}.
In the following sections, I discuss and propose new dimensions that require further development, expanding upon the open questions that I raised in Section 2.

\subsection{Transitioning from Analytics-Heavy to Engagement-Savvy}
The prevailing emphasis on the analytical aspect of data visualization often overshadows other crucial dimensions such as communication, engagement, emotional resonance, and community empowerment \cite{baumer2022course}. While the analytical aspect is vital, it is equally important to foster connections with people, to draw their attention and to make data relatable and actionable for them. By focusing solely on analysis, we risk alienating the very people who could benefit most from the insights data provides. 
Additionally, it is imperative to consider how factors such as educational background, political affiliation, and personal experience shape attitudes and trust in data visualization, particularly among underrepresented populations\cite{peck2019data}.
Incorporating emotional engagement and community-building elements can transform data visualization into a powerful tool for education, advocacy, and collective action, especially on pressing issues such as climate change and public health. It is time to broaden our approach to data visualization, recognizing that its power lies not just in what it reveals, but in how it connects and mobilizes people.

Many open questions remain including: How can we measure the impact of  engagement in data visualizations on public understanding and behavior change given the subjectivity of it ?
What are the best practices for designing visualizations that cater to diverse audiences with varying levels of data, visual and language literacy?
How can community feedback be effectively integrated into the design process of visualizations to enhance relevance and engagement? and
What role do emotions and personal relevance play in the effectiveness of data visualizations, and how can these elements be integrated into design practices?

While we introduced an interdisciplinary approach by borrowing Bloom's Taxonomy from the field of education to more meaningfully measure engagement \cite{mahyar2015towards}, future studies are needed to develop frameworks for identifying factors that lead to disengagement, as well as methods and processes for evaluating and refining engagement practices in data visualization.

\subsection{Democratizing Data \& Broadening Participation}
Effective visualization strategies should aim to democratize access to information by making complex data accessible and understandable to a broad audience \cite{burns2023we}. This involves not only designing comprehensible and navigable visualization and visual interfaces but also incorporating techniques such as storytelling and narrative techniques to convey data-driven insights effectively. By empowering communities with the tools and knowledge to interpret and act upon data, visualization can foster informed decision-making and active participation in addressing societal challenges \cite{baumer2022course}.

Advancing the field of data visualization requires a commitment to inclusivity and accessibility. It is notable that broadening participation includes not only novices and the general public, but also domain experts and decision makers. Our research shows that even these professionals can feel like visualization novices due to a lack of visual analytics expertise \cite{mahyar2019civic, jasim2021communitypulse}. They often request simple visualization techniques and interactive modes that allow them to operate the system independently, without needing to delegate data analysis to analysts and data scientists.
Visualization needs to address the diverse needs and capabilities of various stakeholders by embracing new theoretical frameworks and methodologies. By doing so, we can create visualizations that engage and empower stakeholders, enhance public understanding, and drive positive social change.

Future work should explore questions such as: What are the most effective strategies for designing visualizations that are both comprehensible and engaging to diverse audiences? In "Of Course It's Political" \cite{baumer2022course}, we proposed five new dimensions to help designers and researchers consider their design choices more diligently. However, the paper was just the beginning, sparking a new philosophy. Further research is needed to design and evaluate these techniques for more concrete answers. Other questions include: 
How can storytelling and narrative techniques be optimized to convey complex data in a way that resonates with different demographic groups?
What are the barriers to data literacy among different communities, and how can visualization design help overcome these barriers? and
How can participatory design processes be implemented to ensure that visualizations meet the needs of all stakeholders, especially marginalized groups?

\subsection{Promoting Emotional Resonance \& Inclusive Dialogue}
Effective communication of environmental data requires consideration of the audience's emotional responses. Visualizations that evoke emotion or a sense of urgency can be more impactful than those that present data neutrally. For instance, interactive tools that allow users to visually see the future impacts of climate change on their own neighborhoods can elicit stronger emotional responses and drive action \cite{yavo2023building}. Some effective ways to communicate climate change data to the public include leveraging public art and augmented reality. Public art, with its engaging impact and universal accessibility, helps with raising awareness and fostering public dialogue. 

For instance, RisingEMOTIONS is a data physicalization and public art project that we installed in East Boston in 2020 \cite{aragon2021risingemotions}. Our goal was to engage communities affected by sea-level rise in planning adaptation strategies and increase their involvement in these crucial processes.

This installation visually represents local projected flood levels and public emotions towards the threat of sea-level rise. Placing it in front of the East Boston Public Library, a prominent community hub, ensured high visibility and accessibility to all community members. The project included an opening ceremony and remained on display for two weeks.
 The strategic location played a vital role in engaging community members, as it was installed  in an open space. 
Additionally, the installation was designed to be approachable, allowing viewers to engage effortlessly regardless of prior knowledge about public art. The community's engagement with our project demonstrated the potential for public art to create interest and raise awareness of climate change. 

Augmented reality (AR) enhances public understanding and engagement with climate change impacts by overlaying digital information onto the real-world.
Leveraging AR technology, we conducted the first of its kind Communal eXtended Reality (CXR) study aimed at providing immersive experiences to help people visualize the potential impacts of floods and encourage proactive action \cite{yavo2023building}. We engaged 74 community members by inviting them to ride a local shuttle bus, utilizing public transportation as a communal gathering space. Equipped with VR Head-Mounted Displays (HMDs), participants explored their neighborhood's past and future, physically traversing the island to comprehend the effects of climate change over time. Our study revealed several significant advantages of CXR. Firstly, its immersive and embodied nature made climate change more tangible and immediate to participants. Secondly, its situational elements brought the reality of climate impacts closer to their daily surroundings. Thirdly, the communal setting of the experience emphasized community resilience and responses. A striking observation from this work was how the community aspect transformed negative emotions, such as hopelessness, into actionable plans. This work led to the development of a resiliency plan for the neighborhood. This is particularly important because research has shown that the shocking and negative emotions elicited by confronting the harsh realities of climate change often discourage community members from further engagement with the subject.

Questions to further explore include: 
How can visualizations effectively balance accuracy in depicting climate change data with the emotional engagement needed to prompt action without overwhelming or disengaging audiences?
What are the most effective design strategies for creating public art installations that foster inclusive dialogue and community engagement around climate change impacts?
How can augmented reality (AR) applications be optimized to enhance public understanding of climate change impacts while ensuring accessibility and usability across diverse audiences? and
What are the long-term effects of engaging communities through emotionally resonant visualizations, such as public art and AR, on their attitudes and behaviors towards climate change?




\subsection{Drawing from Other Fields to Build New Theories}

Drawing on the fields of citizen sourcing, crowdsourcing, participatory design, and asset-based design provides valuable insights into enhancing data visualization for societal challenges. Citizen sourcing emphasizes engaging communities in data collection and decision-making processes, ensuring local knowledge and priorities are integrated into sustainability initiatives. Crowdsourcing extends this concept by leveraging collective intelligence to address complex problems through collaborative data analysis and visualization. Participatory design emphasizes the active involvement of stakeholders in the design process, ensuring that visualizations are relevant, accessible, and actionable for diverse audiences.

Platforms that allow community members to contribute data and visualize their findings foster greater public engagement and ownership of sustainability initiatives \cite{mahyar2018communitycrit, jasim2021communityclick}.  For instance, \textbf{CommunityCrit}'s micro-activity workflow proved to be successful in providing a complementary avenue and a new opportunity for community members and those who are not usually engaged (e.g., family with kids, working professionals) to provide meaningful feedback on urban design in a short amount of time \cite{mahyar2018communitycrit}. When deployed in San Diego, CommunityCrit surpassed the current community engagement approach by gathering 352 comments within four weeks of deployment. 
Another example is visualizing crowd-sourced data on local wildlife sightings to track biodiversity and inform conservation efforts. These grassroots visualizations not only democratize data, but also enhance the credibility and reach of sustainability campaigns \cite{english2018crowdsourcing}.
Another example is Chemicals in the Creek, a community-based situated data physicalization to enhance community engagement with open government data \cite{perovich2020chemicals}.

Asset-based design approaches represent a crucial method for communicating complex climate change data to the public \cite{wong2020culture, turner2000asset}. This approach shifts focus to community strengths and resources, utilizing local assets to promote sustainable development and resilience planning. Integrating principles, theories, and frameworks from the aforementioned fields into data visualization practices can result in more inclusive, effective, and empowering visualizations. Such approaches not only enhance communication and understanding but also foster a sense of ownership and collective responsibility toward sustainability challenges.

\section{Conclusion}
In conclusion, reimagining the field of information visualization to address sustainability goals requires a multifaceted approach that prioritizes community engagement. By integrating emotional resonance, inclusivity, and participatory frameworks, we can create visualizations that not only present data but also connect with broad audiences on a deeper level. This connection fosters informed dialogue, empowers communities to influence policies, and drives collective action. Additionally, understanding the epistemology of data visualization is crucial. This involves recognizing how information is constructed, processed, selected, validated, and communicated through visual means, which shapes our perception and interaction with data. By reflecting on the historical and philosophical dimensions of visualization, we can better address its limitations and opportunities for further expansions. Emphasizing the design and development of visualizations with target audiences in mind, and drawing insights from interdisciplinary research, will ensure that visualization is effective in bridging gaps, promoting understanding, and facilitating meaningful engagement. Ultimately, this modern, integrated perspective on data visualization can serve as a powerful catalyst for social empowerment and positive change, paving the way for a more sustainable and resilient future.

\balance
\bibliographystyle{abbrv-doi}

\bibliography{template}
\end{document}